\documentclass[prb,aps,showpacs,twocolumn,superscriptaddress,nobibnotes,epsf]{revtex4}

\usepackage{graphicx}% Include figure files
\usepackage{dcolumn}% Align table columns on decimal point
\usepackage{bm}% bold math
\usepackage{SIunits}
\usepackage{textcomp}
\usepackage{tabularx,booktabs }
\usepackage{threeparttable}

\begin{document}

\title{Superconductivity and Phase Diagram in (Li$_{0.8}$Fe$_{0.2}$)OHFeSe$_{1-x}$S$_x$}

\author{X. F. Lu}
\affiliation{Hefei National Laboratory for Physical Science at
Microscale and Department of Physics, University of Science and
Technology of China, Hefei, Anhui 230026, People's Republic of
China}
\affiliation{Key Laboratory of Strongly-coupled Quantum Matter
Physics, University of Science and Technology of China, Chinese
Academy of Sciences, Hefei, Anhui 230026, People's Republic of
China}

\author{N. Z. Wang}
\affiliation{Hefei National Laboratory for Physical Science at
Microscale and Department of Physics, University of Science and
Technology of China, Hefei, Anhui 230026, People's Republic of
China}
\affiliation{Key Laboratory of Strongly-coupled Quantum Matter
Physics, University of Science and Technology of China, Chinese
Academy of Sciences, Hefei, Anhui 230026, People's Republic of
China}

\author{X. G. Luo}
\affiliation{Hefei National Laboratory for Physical Science at
Microscale and Department of Physics, University of Science and
Technology of China, Hefei, Anhui 230026, People's Republic of
China}
\affiliation{Key Laboratory of Strongly-coupled Quantum Matter
Physics, University of Science and Technology of China, Chinese
Academy of Sciences, Hefei, Anhui 230026, People's Republic of
China}
\affiliation{Collaborative Innovation Center of Advanced Microstructures, Nanjing University, Nanjing 210093, People's Republic of
China}

\author{G. H. Zhang}
\affiliation{CAS Key Laboratory of Materials for Energy Conversion, Shanghai Institute of Ceramics, Chinese Academy of Sciences, Shanghai 200050, People's Republic of
China}
\affiliation{Beijing National Laboratory for Molecular Sciences and State Key Laboratory of Rare Earth Materials Chemistry and Applications,
College of Chemistry and Molecular Engineering, Peking University, Beijing 100871, People's Republic of
China}

\author{X. L. Gong}
\affiliation{CAS Key Laboratory of Mechanical Behavior and Design of Materials, Department of Modern Mechanics, University of Science and
Technology of China, Hefei, Anhui 230027, People's Republic of
China}

\author{F. Q. Huang}
\affiliation{CAS Key Laboratory of Materials for Energy Conversion, Shanghai Institute of Ceramics, Chinese Academy of Sciences, Shanghai 200050, People's Republic of
China}
\affiliation{Beijing National Laboratory for Molecular Sciences and State Key Laboratory of Rare Earth Materials Chemistry and Applications,
College of Chemistry and Molecular Engineering, Peking University, Beijing 100871, People's Republic of
China}

\author{X. H. Chen}
\altaffiliation{Corresponding author} \email{chenxh@ustc.edu.cn}
\affiliation{Hefei National Laboratory for Physical Science at
Microscale and Department of Physics, University of Science and
Technology of China, Hefei, Anhui 230026, People's Republic of
China}
\affiliation{Key Laboratory of Strongly-coupled Quantum Matter
Physics, University of Science and Technology of China, Chinese
Academy of Sciences, Hefei, Anhui 230026, People's Republic of
China}
\affiliation{Collaborative Innovation Center of Advanced Microstructures, Nanjing University, Nanjing 210093, People's Republic of
China}

\begin{abstract}
A series of (Li$_{0.8}$Fe$_{0.2}$)OHFeSe$_{1-x}$S$_x$ (0 $\leq$ x $\leq$ 1) samples were successfully synthesized via hydrothermal reaction method and the phase diagram is established. Magnetic susceptibility suggests that an antiferromagnetism arising from (Li$_{0.8}$Fe$_{0.2}$)OH layers coexists with superconductivity, and the antiferromagnetic transition temperature nearly remains constant for various S doping levels. In addition, the lattice parameters of the both \emph{a} and \emph{c} axes decrease and the superconducting transition temperature T$_c$ is gradually suppressed with the substitution of S for Se, and eventually superconductivity vanishes at $x$ = 0.90. The decrease of T$_c$ could be attributed to the effect of chemical pressure induced by the smaller ionic size of S relative to that of Se, being consistent with the effect of hydrostatic pressure on (Li$_{0.8}$Fe$_{0.2}$)OHFeSe. But the detailed investigation on the relationships between $T_{\rm c}$ and the crystallographic facts suggests a very different dependence of $T_{\rm c}$ on anion height from the Fe2 layer or $Ch$-Fe2-$Ch$ angle from those in FeAs-based superconductors.
\end{abstract}


%% 74.70.Xa: Pnictides and chalcogenides; 74.25.-q  Properties of superconductors; 74.25.Dw Superconductivity phase diagrams

\vskip 300 pt

\maketitle

Since the discovery of superconductivity in
LaFeAsO$_{1-x}$F$_x$ with T$_c$ $\sim$ 26K \cite{1}, the
iron-based superconductors, as the second family of compounds exhibiting high $T_{\rm c}$ after the cuprates, have attracted wide attention \cite{2,3,4}. $\beta$-FeSe, which shows superconducting transition at $\sim$ 10 K and owns the simplest crystal structure among iron-based
superconductors, is thought to be a promising system to investigate the mechanism
of high $T_{\rm c}$ superconductivity in iron-based superconductors \cite{5}. By applying hydrostatic pressure or
intercalating alkali atoms between FeSe layers (with chemical
formula $A_x$Fe$_{2-y}$Se$_2$ [$A$ = K, Rb, Cs, Tl/K, Tl/Rb,
etc.], the $T_{\rm c}$ could be enhanced to higher than 30 K\cite{6,7,8,9,10}. However, in $A_x$Fe$_{2-y}$Se$_2$, the
obvious phase separation between the superconducting
phase and the inter-grown antiferromagnetic(AFM)
insulating phase with an extremely high N${\rm \acute{e}}$el temperature of
$\sim$ 560K and Fe vacancy ordering \cite{11,
12,13,14}, makes it difficult to study the
underlying physics of FeSe layers. In addition, other FeSe-derived superconductors, such
as alkali metal ions and NH$_3$ molecules or organic-molecules intercalated FeSe\cite{15,16,17,18}, are
extremely air-sensitive, which prevents the further investigation of their physical properties. Thus, it is urgent to find other
FeSe-derived superconductors with new spacer layers.

In iron-based superconductors, both carrier doping and isovalent
substitution can tune the superconducting properties \cite{4,19}.
Similar to the external pressure effect, isovalent substitution
would not change carrier density but could introduce or enhance
superconductivity, as found in the iron arsenides \cite{19,20}. For instance, through
substituting As with isovalent P, bulk superconductivity emerges in
LaFeAs$_{1-x}$P$_x$O with $T_{\rm c}$ of 10.8 K, which is understood in
terms of chemical pressure and bond covalency \cite{20}. However, in
FeSe-derived superconductors K$_x$Fe$_{2-y}$Se$_{2-z}$S$_z$, the $T_{\rm c}$ is suppressed with S substituting for Se, and goes to
zero at 80\% of S, which has been attributed to the increase of
Fe-Se tetrahedron irregularity and Fe1 site occupancy \cite{21}.
Recently, an air-stable FeSe-derived superconductor
(Li$_{0.8}$Fe$_{0.2}$)OHFeSe was reported with $T_{\rm c}$ of $\sim$ 40 K and the precise crystal structure has been unambiguously determined \cite{22,23}. Moreover, there exists a canted AFM order originating
from (Li$_{0.8}$Fe$_{0.2}$)OH layer, which coexists with
superconductivity. In this work, we report on the successful synthesis
of (Li$_{0.8}$Fe$_{0.2}$)OHFeSe$_{1-x}$S$_x$ (0 $\leq$ x $\leq$ 1) by using
hydrothermal reaction method. The evolution of superconducting
properties and structure parameters with S content in
(Li$_{0.8}$Fe$_{0.2}$)OHFeSe$_{1-x}$S$_x$ are investigated. The
results reveal that both a- and c-axis lattice parameters decrease
almost linearly with the increase of S content. Superconductivity is
suppressed by the substitution of S for Se, and finally vanishes at
$x$ = 0.90. Moreover, the AFM order locating within
the (Li$_{0.8}$Fe$_{0.2}$)OH layer coexists with superconductivity, and the AFM transition temperature almost remains unchanged with S
content.

A series of (Li$_{0.8}$Fe$_{0.2}$)OHFeSe$_{1-x}$S$_x$ samples with nominal composition $x$ = 0.0-1.0 were synthesized by
hydrothermal reaction method, as described in the previous report \cite{22,23}. First, in order to ensure the reagents were fully dissolved and mixed, 0.012-0.02 mol selenourea (Alfa Aesar, 99.97\% purity) and sulfourea (Sinopharm Chemical Reagent, A.R.purity) were stoichiometrically weighted, dissolved in 10 ml water, and stirred for 10-20 minutes in the Teflon-lined autoclave. Then 0.0075 mol Fe powder (Aladdin Industrial, A.R.purity) and 12 g LiOH$\cdot$H$_2$O (Sinopharm Chemical Reagent, A.R.purity) were thrown into the autoclave and mixed. Finally, the Teflon-lined autoclave was tightly sealed and heated at 150-160 $\celsius$ for 3-10 days. The polycrystalline samples acquired from the reaction systems were washed with deionized water repeatedly, and dried at room temperature.

Powder x-ray diffraction(XRD) data of samples were collected by
using x-ray diffractometer (SmartLab-9, Rikagu Corp.) with Cu
K$\alpha$ radiation and a fixed graphite monochromator in the
2-$\theta$ range of 5${\rm ^\circ}$-70 ${\rm ^\circ}$ at room temperature. The average stoichiometries of
Fe, Se and S of the polycrystalline samples were determined from
energy-dispersive x-ray spectroscopy (EDX) analysis. The actual S contents $x$
were determined by EDX to be 0, 0.08, 0.16, 0.22, 0.28, 0.41, 0.53, 0.66, 0.77, 0.90, and 1.0 for the 11 samples used in this work with the nominal molar reagents ratios of Sulfourea/(Sulfourea+Selenourea)= 0, 0.1, 0.2, 0.3, 0.4, 0.5, 0.6, 0.7, 0.8, 0.9, and 1.0, respectively. Magnetization measurements were carried out on SQUID MPMS-XL5 (Quantum Design). Refinements of the
XRD data were performed by using GSAS software \cite{24,25}.

\begin{figure}[htp]
\centering
\includegraphics[width=0.47\textwidth]{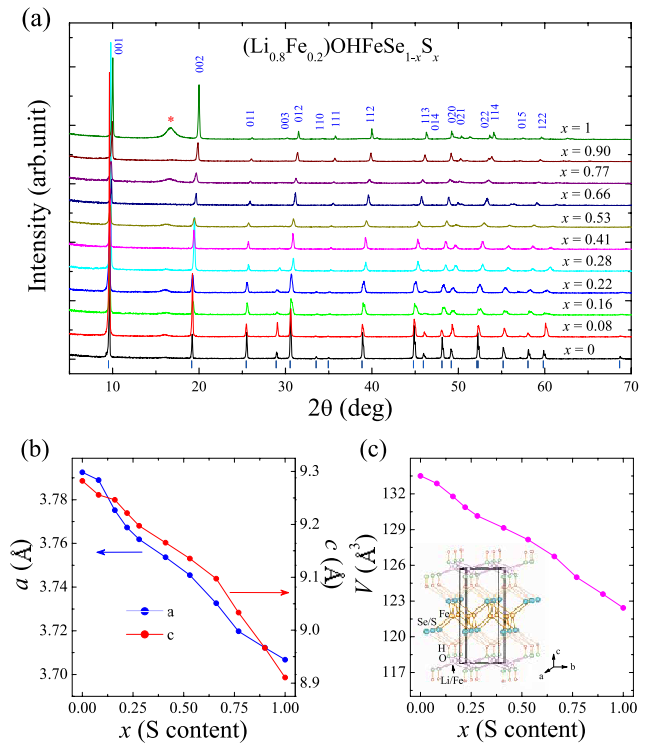}
\caption{(Color online) (a) The powder XRD patterns of
(Li$_{0.8}$Fe$_{0.2}$)OHFeSe$_{1-x}$S$_x$ (0 $\leq$ x $\leq$ 1) at room
temperature. The asterisk represents the nanoscale FeS. (b) and (c) The lattice parameters of the \emph{a} and \emph{c} axes
and unit cell volume as a function of the S content. The inset figure in (b) is
the crystal structure of (Li$_{0.8}$Fe$_{0.2}$)OHFeSe$_{1-x}$S$_x$\cite{23}. }
\end{figure}

Powder XRD patterns of (Li$_{0.8}$Fe$_{0.2}$)OHFeSe$_{1-x}$S$_x$ samples are shown in Fig.1(a), all of which were collected
at room temperature. The XRD patterns of (Li$_{0.8}$Fe$_{0.2}$)OHFeSe$_{1-x}$S$_x$ are similar to that of
(Li$_{0.8}$Fe$_{0.2}$)OHFeSe and all reflections can be well indexed by the
tetragonal structure on the basis of the space group of $P$4/$nmm$
(No. 129), except for the broad one at about 2$\theta$ = 16 {\rm $^\circ$}. The broad reflection at 2$\theta \approx$  16 ${\rm ^\circ}$ may be attributed to the nanoscale FeS produced in the
low-temperature synthesis procedure \cite{26}. As shown in
Fig.1(a), all reflections shift to the higher 2$\theta$ side with the increase of
S content. Figure 1(b) shows the evolution of the lattice parameters
along the \emph{a} and \emph{c} axes as a function of S content $x$. With
increasing $x$, the lattice parameters along both the \emph{a} and \emph{c} axes monotonically decrease, indicating the
lattice contraction with increasing S content, which is consistent with the relatively smaller ionic size of S$^{2-}$ compared with
Se$^{2-}$. As a result, the unit cell volume $V$ = $a\times a\times
c$ also decreases monotonically. The lattice shrinking progressively
with S substitution is consistent with Vegard's Law, which is
similar to K$_x$Fe$_{2-y}$Se$_{2-z}$S$_z$ and FeSe$_{1-x}$S$_x$ \cite{21, 27}.

\begin{figure}[htp]
\centering
\includegraphics[width=0.5\textwidth]{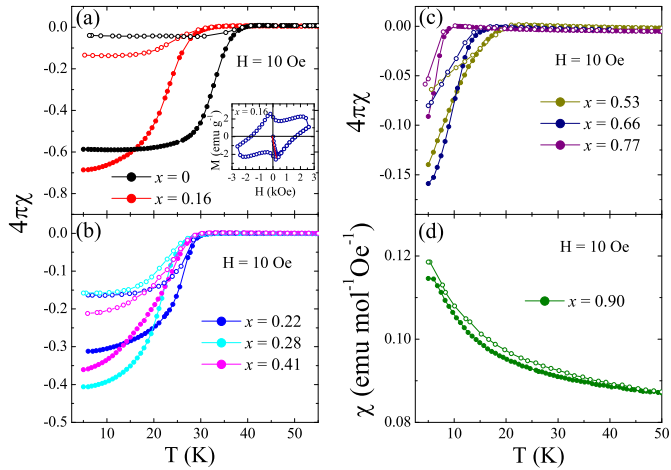}
\caption{(Color online) Temperature dependence of the dc magnetic susceptibility of the as-synthesized
samples (Li$_{0.8}$Fe$_{0.2}$)OHFeSe$_{1-x}$S$_x$, taken in zero-field cooling (ZFC) (solid symbols) and field
cooling (FC) (open symbols) modes under an external field of 10
Oe. The inset of (a) is the $M$-$H$ loop of
(Li$_{0.8}$Fe$_{0.2}$)OHFeSe$_{1-x}$S$_x$ ($x$ = 0.16) taken at 5 K.}
\end{figure}

Figure 2 shows the temperature dependence of magnetic susceptibility
$\chi$ for the superconducting samples under a magnetic field
of 10 Oe. $T_{\rm c}$ determined from zero-field-cooling (ZFC) magnetic susceptibility
shifts gradually to low temperature as the S content increases. When
S content increases up to $x$ = 0.90, no diamagnetic signal can
be observed above 5 K and the temperature dependence of magnetic
susceptibility shows paramagnetic behavior. Additionally, the
shielding fractions at 5 K of (Li$_{0.8}$Fe$_{0.2}$)OHFeSe$_{1-x}$S$_x$
($x$ = 0.16) estimated from the ZFC curves is 69\%, suggesting a bulk
superconductivity at 37 K. The $M$-$H$ loop of $x$ = 0.16 sample
measured at 5 K is presented in the inset of Fig. 2(a). A linear-$H$ dependence
of diamagnetic magnetization with negative slope can be observed up
to $\sim$ 150 Oe, which is in accordance with the superconducting
transition observed in the temperature dependence of susceptibility.
According to Mizuguchi's report \cite{27}, the S substitution in
FeSe can stabilize the superconducting state. However, in our case,
the sizes of crystalline grains from a low-temperature solution
synthetic method are usually small and reduce the superconductive
shielding fraction of samples, especially when $x$
exceeds 0.50.

\begin{figure}[htp]
\centering
\includegraphics[width=0.5\textwidth]{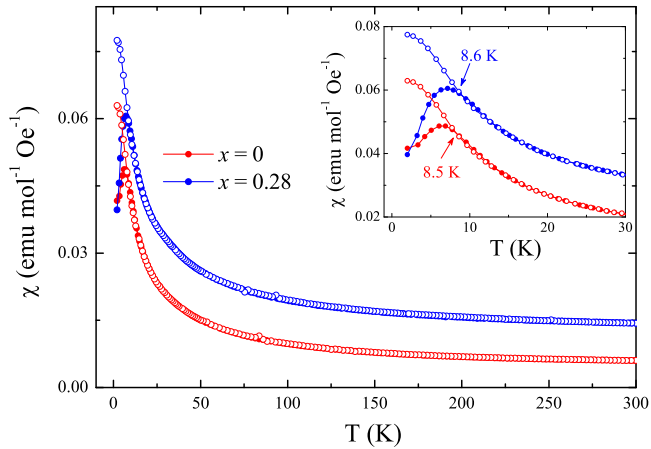}
\caption{Color online) The temperature dependence of magnetic susceptibility $\chi$ for samples from 2 to 300 K by applying an external field of 1 T.}
\end{figure}

Figure 3 shows the temperature dependence of magnetic susceptibility
$\chi$ for $x$ = 0 and 0.28 samples from 2 to 300 K by applying an
external field of 1 T. The superconductivity seemed to be suppressed
under this field. However, the magnetic order in the
(Li$_{0.8}$Fe$_{0.2}$)OH layer creates an internal field and completely
suppresses the Meissner effect under 1 T. Thus, there is no
diamagnetic signal observed under 1 T. Moreover, the temperature
dependence of magnetization displays a Curie-Weiss behavior above 10
K. A sudden decrease in the $\chi$ happens in the ZFC curve around 8
K for both of the samples with $x$ = 0 and 0.28. FC
and ZFC magnetic susceptibilities bifurcate for both samples at about 8 K. The bifurcation is quite weak, suggesting a
weak ferromagnetic component due to a possible canted antiferromagnetic
order, as derived from nuclear magnetic resonance (NMR) measurements
in other work \cite{28}. The temperatures corresponding to the maximum
of the ZFC susceptibility and the bifurcating temperature of ZFC and
FC susceptibilities for the sample with $x$ = 0.28 are almost the
same as those observed in the S-free sample, strongly suggesting
that this magnetic order is formed within the
(Li$_{0.8}$Fe$_{0.2}$)OH layers, so that the substitution of S for
Se cannot affect the magnetic transition. This is consistent with
the NMR results in our other work, indicating that this magnetic
ordered state originated from the (Li$_{0.8}$Fe$_{0.2}$)OH
layers\cite{28}.

\begin{figure}[htp]
\centering
\includegraphics[width=0.5\textwidth]{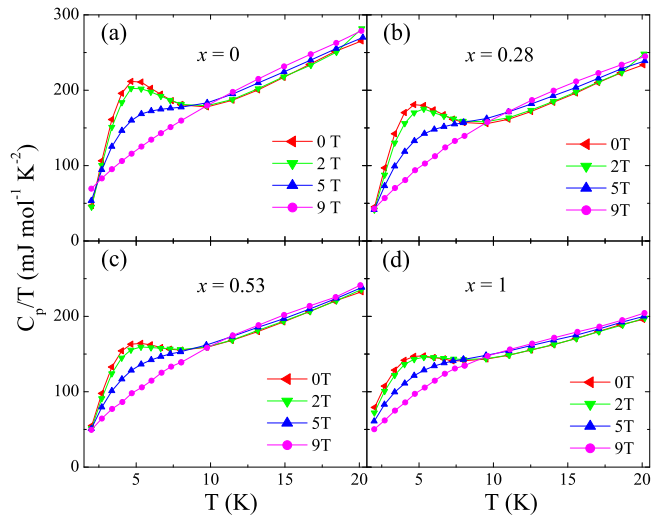}
\caption{Color online) The specific heat of (Li$_{0.8}$Fe$_{0.2}$)OHFeSe$_{1-x}$S$_x$ ($x$ = 0,0.28,0.53,1.0) under different external fields}
\end{figure}

In order to confirm the AFM transition, we performed
thermodynamic measurements. Figure 4 shows the specific heat
measured under different magnetic fields. The specific heat for all the
samples with different S contents begins to rise at about 8 K, which
is consistent with the anomaly temperature in the magnetic
susceptibility. Such rise is suppressed with increasing magnetic
fields and becomes very obscure as the field increases up to 9 T.
Surprisingly, the temperature for the maximum of specific heat
remains unshifted at 5 K in various magnetic fields. These
features are consistent with the antiferromagnetic order proposed
above. These results further suggest that the AFM ordering should arise from the (Li$_{0.8}$Fe$_{0.2}$)OH layer.

Based on the magnetic measurements displayed in Figs. 2 and 3 as well as the
thermodynamic results shown in Fig.4, the phase diagram is mapped
out for the (Li$_{0.8}$Fe$_{0.2}$)OHFeSe$_{1-x}$S$_x$ (0 $\leq  x \leq$ 1), as
shown in Fig. 5, where $T_{\rm c}$ is determined by susceptibility
and magnetic transition temperature is determined by the specific
heat. The  $T_{\rm c}$ gradually decreases and vanishes at $x$ =
0.90, although the substituted S is isovalent to Se. The decrease of
$T_{\rm c}$ is accompanied by the reduction of the a- and c-axis
lattice parameters, suggesting the suppression effect of the
chemical pressure on $T_{\rm c}$. This is in accordance with the
suppression effect of external pressure on superconductivity in
(Li$_{0.8}$Fe$_{0.2}$)OHFeSe.

\begin{figure}[htp]
\centering
\includegraphics[width=0.5\textwidth]{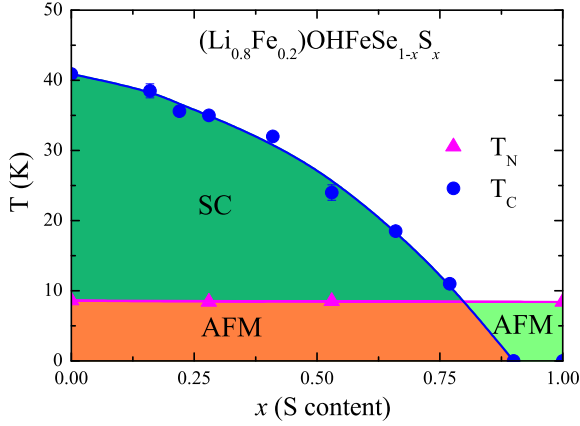}
\caption{Color online) The phase diagram of (Li$_{0.8}$Fe$_{0.2}$)OHFeSe$_{1-x}$S$_x$ derived from the magnetic susceptibility. The solid lines are a guide for the eye.}
\end{figure}

\begin{figure}[htp]
\centering
\includegraphics[width=0.5\textwidth]{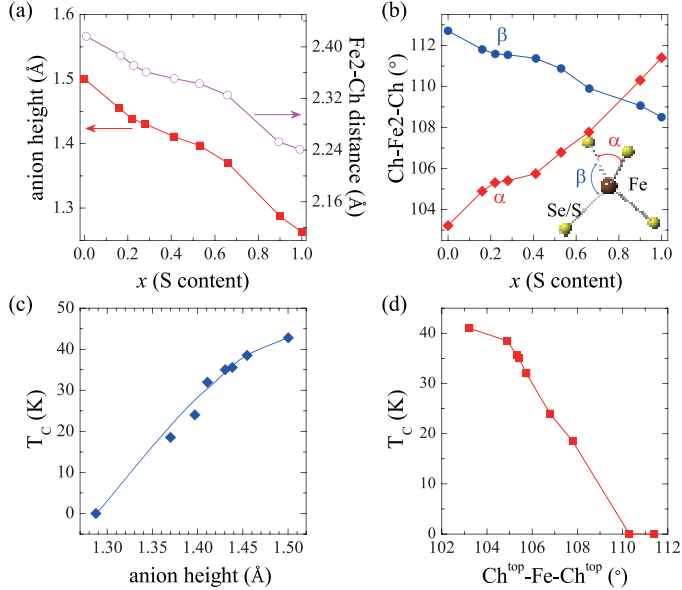}
\caption{Color online) (a) and (b) The evolution of $Ch$-Fe2-$Ch$ angles(2$\times$ and 4$\times$)
and Fe2-$Ch$ bond lengths in the Fe2-$Ch$ tetrahedron and the anion height
from Fe2 layer with S substitution, where $Ch$ is the chalcogen S and Se. (c) $T_{\rm c}$ plotted against chalcogen anion
height for (Li$_{0.8}$Fe$_{0.2}$)OHFeSe$_{1-x}$S$_x$ (0 $\leq x \leq$ 1) samples. (d) The relationship between
$T_{\rm c}$ and Ch-Fe2-Ch angles(2$\times$).}
\end{figure}

It is believed that there are close relationships between $T_{\rm c}$ and crystallographic details. In our case, there is no other Fe site between the
(Li$_{0.8}$Fe$_{0.2}$)OH layer and the FeSe$_{1-x}$S$_x$ layer, and both the
chalcogen ($Ch$) anion height from the Fe2 layer and the Fe2-$Ch$ bond distance in FeSe$_{1-x}$S$_x$ slab monotonically decrease with increasing S
content, as shown in Fig. 6(a). The evolution of the $Ch$-Fe2-$Ch$ angle in the Fe2-$Ch$
tetrahedron is shown in Fig. 6(b). The $Ch$-Fe2-$Ch$ angles change towards the ideal value of a regular tetrahedron (109.47${\rm ^\circ}$). According to a previous report \cite{29}, $T_{\rm c}$ is closely
connected to the anion height from Fe layer ($h$) and a maximum
$T_{\rm c}$ could be achieved with $h_0\approx$1.38 \AA for
FeAs-derived superconductors. For the FeSe-derived superconductors,
the anion height dependence of $T_{\rm c}$ has been established as a
\textsf{V}-shape\cite{22} with a minimum $T_{\rm c}$ at $h$ = 1.45 \AA \cite{22}, distinct from the inverse
\textsf{V}-shape one in FeAs-based superconductors. Both S and Te
substitutions for Se in FeSe would enhance $T_{\rm c}$ \cite{27,
30}, which could be attributed to the variation of anion height and
follow the law of \textsf{V}-shape dependence of $T_{\rm c}$.
However, in regard to K$_x$Fe$_{2-y}$Se$_{2-z}$S$_z$, the anion
height dependence of $T_{\rm c}$ violates this law, which can be
explained in terms of the existence of the Fe vacancies, which
results in a nonmonotonic change of the anion height with S content
\cite{21}. In Fig. 6(c), we plot $T_{\rm c}$ as a function of chalcogen
height from the Fe layer for (Li$_{0.8}$Fe$_{0.2}$)OHFeSe$_{1-x}$S$_x$, which shows that the
$T_{\rm c}$ is monotonically decreased with shrinking of the anion
height from Fe layer, with no sudden change in slope at $h$=1.38 or 1.45 \AA. This does not follow the previous \textsf{V}-shape in FeSe-derived superconductors or the inverse \textsf{V}-shape in FeAs-based superconductors, suggesting the existence of peculiar physics in the (Li$_{0.8}$Fe$_{0.2}$)OHFeSe$_{1-x}$S$_x$ system. In FeAs-based superconductors, it is also thought that the regular tetrahedron of FeAs$_4$ might favor higher $T_{\rm c}$ and this might hold in FeSe-derived superconductors. However, Fig. 6(d) shows that $T_{\rm c}$ decreases monotonically as the $Ch$-Fe2-$Ch$ angle goes to the ideal value of a regular tetrahedron, implying that a tetrahedron distortion in FeSe-derived superconductors may promote the superconductivity.

Another intriguing phenomenon of the phase diagram shown in Fig. 5
is that although the $T_{\rm c}$ can be effectively suppressed by S
substitution, the AFM transition temperature remains almost
unchanged. In the superconducting region of S content, AFM ordering
exists deeply inside the superconducting state and coexists with
superconductivity in the whole region, but seems to have no
connection with the superconductivity. For a conventional
superconductor, local magnetic moments or magnetic order is usually
unfavorable to superconductivity. However,  AFM order from (Li$_{0.8}$Fe$_{0.2}$)OH
layers seems not to affect superconductivity occurring in the
conducting FeSe layers for (Li$_{0.8}$Fe$_{0.2}$)OHFeSe$_{1-x}$S$_x$. Actually,
it is found in FeAs-based superconductors that magnetism or magnetic
moments outside the conducting FeAs layers can have negligible
suppression effect on superconductivity. In
Eu$_{1-x}$La$_x$Fe$_2$As$_2$, AFM can also exist deep inside the
superconducting region with both $T_{\rm c}$ and AFM transition
temperature increasing with enhancing external
pressure\cite{31}. Replacement of magnetic Nd, Pr, Sm, and Gd
for nonmagnetic La in LaFeAsO$_{1-x}$F$_{x}$ or LaFeAsO$_{1-\delta}$
can enhance $T_{\rm c}$ effectively\cite{2,3,RenZA,RenZA1}. These
facts strongly manifest the unconventional superconductivity in the
Fe-based superconductors. It also suggests that the correlation
along the \emph{c} axis plays a trivial role in the superconductivity in
the Fe-based superconductors.

In summary, we successfully synthesized a series of
(Li$_{0.8}$Fe$_{0.2}$)OHFeSe$_{1-x}$S$_x$ (0 $\leq x \leq$ 1) samples through the
hydrothermal method. Due to the smaller ionic size of S relative to
that of Se, the S substitution leads to shrinking of the lattice
parameters both along the \emph{a} axis and the \emph{c} axis. Magnetic
susceptibility and specific heat were also studied. Based on the
magnetic susceptibility and thermodynamic results of all the
samples, the phase diagram of (Li$_{0.8}$Fe$_{0.2}$)OHFeSe$_{1-x}$S$_x$ is
mapped out. $T_{\rm c}$ is suppressed from 40 K to zero as S content
increases from 0 to 0.90. The effect of chemical pressure resulting
from S substitution for Se is considered as a possible mechanism of
the suppression of $T_{\rm c}$, which is in agreement with the
effect of external pressure previously investigated in
(Li$_{0.8}$Fe$_{0.2}$)OHFeSe. But the relationships between $T_{\rm c}$ and the crystallographic details reveal that the dependence of $T_{\rm c}$ on anion height from the Fe2 layer or the $Ch$-Fe2-$Ch$ angle is distinct from those summarized in FeAs-based superconductors. Magnetic susceptibility at 1 T and the specific heat suggest that an AFM transition around 8 K originates from (Li$_{0.8}$Fe$_{0.2}$)OH layers. The magnetic transition temperature does not
alter with S concentration, and superconductivity coexists with antiferromagnetism in the superconducting region of S content.

This work is supported by the National Natural Science Foundation of
China (China) (Grants No. 11190021, No. 11174266, and No. U1330105), the "Strategic
Priority Research Program(B)" of the Chinese Academy of Sciences (China)
(Grant No. XDB04040100), the National Basic Research Program of
China (China) (973 Program, Grants No. 2012CB922002 and No. 2011CBA00101), National Key Project of Magneto-Restriction Fusion Energy Development Program (Grant No. 2013GB110002), and the Chinese Academy of Sciences.

\end{document}